# Heterogeneous item populations across individuals: Consequences for the factor model, item inter-correlations, and scale validity

André Beauducel* & Norbert Hilger

*Institute of Psychology, University of Bonn, Germany*

## Abstract

The paper is devoted to the consequences of blind random selection of items from different item populations that might be based on completely uncorrelated factors for item inter-correlations and corresponding factor loadings. Based on the model of essentially parallel measurements, we explore the consequences of presenting items from different populations across individuals and items from identical populations within each individual for the factor model and item inter-correlations in the total population of individuals. Moreover, we explore the consequences of presenting items from different as well as identical item populations across and within individuals. We show that correlations can be substantial in the total population of individuals even when -in subpopulations of individuals- items are drawn from populations with uncorrelated factors. In order to address this challenge for the validity of a scale, we propose a method that helps to detect whether item inter-correlations result from different item populations in different subpopulations of individuals and evaluate the method by means of a simulation study. Based on the analytical results and on results from a simulation study, we provide recommendations for the detection of subpopulations of individuals responding to items from different item populations.

* Corresponding author:

André Beauducel, Institute of Psychology, Rheinische Friedrich-Wilhelms-Universität Bonn,

Kaiser-Karl-Ring 9, 53111 Bonn, Germany. Email: beauducel@uni-bonn.de

Sometimes participants work on different items which are randomly selected from an item pool. This is especially not a problem when a model of parallel measurements holds so that the items measure exactly the same common factor to the same degree. The items are then interchangeable so that each individual may work on different items. The assumption of the equivalence of items within an item bank is usually stated in the context of item response theories (IRT). For example, a one-parameter logistic model (Rasch, 1960) has often been used as a basis for the construction of a set of equivalent items. Whereas a wealth of studies on item calibration and automatic item selection, mainly in the context of the one-parameter logistic model is available (e.g. Hartig, Köhler & Naumann, 2020; van Buuren & Eggen, 2017; Wang, Chang & Douglas, 2012; Wyse & McBride, 2021), the consequences of blind random selection of items from different item populations on Pearson correlations and the corresponding factor model have not yet been investigated comprehensively. Moreover, it is often assumed that items are selected from the same item population for each individual. However, it may not always *a priori* be known whether the items in each subpopulation of individuals are drawn from one and the same population of items. The heterogeneity of item populations across subpopulations of individuals may be a challenge for the validity of the scale because it implies that the items do not measure the same construct in each subpopulation of individuals. Due to random item selection, items may have convergent validity in some subpopulations of individuals and they may have discriminant validity in other subpopulations of individuals. Thus, we investigate variations of convergent and discriminant validity (Campbell & Fiske, 1959) across subpopulations of individuals. Based on item-sampling theory (Lord & Novick, 1968), we investigate the consequences of blind random selection of items from different item populations that might be based on uncorrelated factors for item inter-correlations and corresponding factor loadings. Thus, the focus of the present work is not on small shifts in item factor loadings between a small number of subpopulations of individuals. This issue can be addressed by means of investigations of item invariance, for example, by means of multiple-group confirmatory factor analysis (Vandenberg & Lance, 2000). However, Maraun and Heene (2015) have shown that factorial invariance across populations of individuals does not imply that factors in the populations are equivalent. This finding was the starting point for the present investigation of a very large number of subpopulations of individuals that may partially work on items from different populations that are based on completely uncorrelated factors.

Although the items are selected from different item populations, we consider the case that they are arranged as if all individuals worked on one and the same item so that the resulting item scores may be based on heterogeneous item populations across individuals. The model of essentially parallel measurements is a reasonable basis for such an arrangement of data because it may imply that all items are from the population. However, it will be shown in the following that substantial and equal inter-item correlations that are typically regarded as a basis for essentially parallel measurements (Lord & Novick, 1968) do not necessarily imply that different items performed by the individuals measure one and the same common factor. It should be noted, however, that the focus of the present paper is not on unidimensionality of an item population as it is investigated when all individuals work on the same items. Several important advices and methods regarding the investigation of unidimensionality by means of factor models when all individuals work on the same items are meanwhile available (Ferrando & Lorenzo-Seva, 2018; Sellbom & Tellegen, 2019). The present paper is rather devoted to the investigation of the unidimensionality of factor models when not all individuals work on the same items. The measured variables for factor models are often items, so that we use the term items for measured variables in the following. However, it is also possible that scales are used as measured variables. The results of the present investigation may also be relevant when different individuals work on different scales although it is not known which individuals worked on which scale as it might occur in big data sets that are combined from different sources.

First, some definitions are presented, second, we explore the consequences of presenting items from different item populations across individuals and items from identical item populations within each individual for item inter-correlations and the corresponding factor model. Third, we explore the consequences of presenting items from different as well as identical item populations across and within individuals for the factor model. Fourth, we develop a method that helps to identify whether item inter-correlations result from different item populations and evaluate the method by means of a simulation study. Finally, the results and conclusions are summarized and some recommendations for further research are given.

## Definitions

Consider two populations of items $\mathbf{X}_{s1}$ and $\mathbf{X}_{s2}$ with each item population containing $p$ random variables representing the item scores in the population of individuals. Each item population is based on another uncorrelated common factor. This can be written as

$$\mathbf{X}_{s1i} = \mathbf{\Lambda f}_1 + \mathbf{\Psi e}_{1i} + \boldsymbol{\mu}_{1i} \text{ and } \mathbf{X}_{s2i} = \mathbf{\Lambda f}_2 + \mathbf{\Psi e}_{2i} + \boldsymbol{\mu}_{2i}, \text{ for } i = 1 \text{ to } p, \quad (1)$$

where $\mathbf{f}_1$ and $\mathbf{f}_2$ are common factors with $E(\mathbf{f}_1) = 0, E(\mathbf{f}_2) = 0, E(\mathbf{f}_1\mathbf{f}_1') = 1, E(\mathbf{f}_2\mathbf{f}_2') = 1, E(\mathbf{f}_1\mathbf{f}_2') = 0$, the $p \times 1$ matrix of common factor loadings $\mathbf{\Lambda}$, the uncorrelated unique factors $\mathbf{e}_1$ and $\mathbf{e}_2$, with $E(\mathbf{e}_1) = \mathbf{0}, E(\mathbf{e}_2) = \mathbf{0}, E(\mathbf{e}_1\mathbf{e}_1') = diag(E(\mathbf{e}_1\mathbf{e}_1')) = \mathbf{I}, E(\mathbf{e}_2\mathbf{e}_2') = diag(E(\mathbf{e}_2\mathbf{e}_2')) = \mathbf{I}$, and $E(\mathbf{e}_1\mathbf{e}_2') = \mathbf{0}$, $E(\mathbf{f}_1\mathbf{e}_1') = \mathbf{0}$, $E(\mathbf{f}_1\mathbf{e}_2') = \mathbf{0}, E(\mathbf{f}_2\mathbf{e}_2') = \mathbf{0}, E(\mathbf{f}_2\mathbf{e}_1') = \mathbf{0}$, the positive definite, diagonal unique factor loading matrix $\mathbf{\Psi}$, and the vectors of item means $\boldsymbol{\mu}_{1i}$ and $\boldsymbol{\mu}_{2i}$. A model of essentially parallel measurements is assumed for each item population so that all elements in $\mathbf{\Lambda}$ are equal. It is also assumed that $\mathbf{\Lambda}$ and $\mathbf{\Psi}$ are identical in both item populations. The expectation of the item means for the two item populations in the population of individuals is $E(\boldsymbol{\mu}_{1i}) = \mathbf{0}$ and $E(\boldsymbol{\mu}_{2i}) = \mathbf{0}$, the variability of the item means within each item population is $E(\boldsymbol{\mu}_{1i}\boldsymbol{\mu}_{1i}') = \mathbf{\Omega}_1 = diag(\mathbf{\Omega}_1)$ and $E(\boldsymbol{\mu}_{2i}\boldsymbol{\mu}_{2i}') = \mathbf{\Omega}_2 = diag(\mathbf{\Omega}_2)$ and $\mathbf{\Omega}_1 = \mathbf{\Omega}_2$. It is assumed that $E(\mathbf{f}_1\boldsymbol{\mu}_{1i}') = \mathbf{0}$, $E(\mathbf{e}_1\boldsymbol{\mu}_{1i}') = \mathbf{0}$, $E(\mathbf{f}_1\boldsymbol{\mu}_{2i}') = \mathbf{0}$, $E(\mathbf{e}_{1i}\boldsymbol{\mu}_{2i}') = \mathbf{0}$, $E(\mathbf{f}_2\boldsymbol{\mu}_{1i}') = \mathbf{0}$, $E(\mathbf{e}_{2i}\boldsymbol{\mu}_{1i}') = \mathbf{0}$, $E(\mathbf{f}_2\boldsymbol{\mu}_{2i}') = \mathbf{0}$, $E(\mathbf{e}_{2i}\boldsymbol{\mu}_{2i}') = \mathbf{0}$.

Let the total population of individuals work on two items $\mathbf{X}_{t1}$ and $\mathbf{X}_{t2}$ that are selected at random from item populations $\mathbf{X}_{s1}$ and $\mathbf{X}_{s2}$. Four item population combinations are possible, the two items may be selected from $\mathbf{X}_{s1}$, the two items may be selected from $\mathbf{X}_{s2}$, the first item may be selected from $\mathbf{X}_{s1}$ and the second item from $\mathbf{X}_{s2}$, or the first item may be selected from $\mathbf{X}_{s2}$ and the second item from $\mathbf{X}_{s1}$. In consequence, there are four subpopulation of individuals, each one responds to one combination of items from the two item populations. The selection of items from the two item populations into $\mathbf{X}_{t1}$ and $\mathbf{X}_{t2}$ can be written as

$$\begin{bmatrix}\mathbf{X}_{t1}\\ \mathbf{X}_{t2}\end{bmatrix}=\begin{bmatrix}\mathbf{X}_{s1},\mathbf{X}_{s1},\mathbf{X}_{s2},\mathbf{X}_{s2}\\ \mathbf{X}_{s1},\mathbf{X}_{s2},\mathbf{X}_{s1},\mathbf{X}_{s2}\end{bmatrix}=\begin{bmatrix}x_{s1j}\cdots,x_{s1j}\cdots,x_{s2j}\cdots,x_{s2j}\cdots\\ x_{s1j}\cdots,x_{s2j}\cdots,x_{s1j}\cdots,x_{s2j}\cdots\end{bmatrix}. \tag{2}$$

**Items from different populations across individuals and from identical populations within individuals**

Although four combinations of the two items are possible, in a first step, the condition that the items are from two item populations of uncorrelated factors and that each individual responds to two items based on the same common factor, is considered. Equation 2 can then be simplified to

$$\begin{bmatrix}\mathbf{X}_{t1}\\ \mathbf{X}_{t2}\end{bmatrix}=\begin{bmatrix}\mathbf{X}_{s1},\mathbf{X}_{s2}\\ \mathbf{X}_{s1},\mathbf{X}_{s2}\end{bmatrix}=\begin{bmatrix}x_{s1j}\cdots,x_{s2j}\cdots\\ x_{s1j}\cdots,x_{s2j}\cdots\end{bmatrix}. \tag{3}$$

The following theorem implies that, when each individual responds to items based on the same factor $\mathbf{f}_1$ or $\mathbf{f}_2$, whereas different individuals respond to items based on the two uncorrelated factors $\mathbf{f}_1$ or $\mathbf{f}_2$, and when the variances of the item means in the two item populations equal the variance of the means in the total item population, the reproduced covariances are the same as when all individuals work on items based on a single factor. The single factor model is defined as $\mathbf{\Lambda f}+\mathbf{\Psi e}+\boldsymbol{\mu}$ with $E(\mathbf{f})=0, E(\mathbf{e})=0,$ $E(\mathbf{ff}')=1,$ $E(\mathbf{ee}')=diag(E(\mathbf{e\,e}'))=\mathbf{I}$, $E(\mathbf{fe}')=\mathbf{0}$, $E(\mathbf{f}\boldsymbol{\mu}')=\mathbf{0}$, $E(\mathbf{e}\boldsymbol{\mu}')=\mathbf{0}$, $E(\boldsymbol{\mu\mu}')=\mathbf{\Omega}=diag(\mathbf{\Omega})$.

**Theorem 1.**

*If* $\mathbf{\Omega}=\begin{bmatrix}\mathbf{\Omega}_1 & \mathbf{0}\\ \mathbf{0} & \mathbf{\Omega}_2\end{bmatrix}$ *then*

$$\mathbf{\Sigma}=E\left\{(\mathbf{\Lambda}[\mathbf{f}_{1i},\mathbf{f}_{2i}]+\mathbf{\Psi}[\mathbf{e}_{1i},\mathbf{e}_{2i}]+[\boldsymbol{\mu}_{1i},\boldsymbol{\mu}_{2i}])(\mathbf{\Lambda}[\mathbf{f}_{1i},\mathbf{f}_{2i}]+\mathbf{\Psi}[\mathbf{e}_{1i},\mathbf{e}_{2i}]+[\boldsymbol{\mu}_{1i},\boldsymbol{\mu}_{2i}])'\right\}=E\left\{(\mathbf{\Lambda f}+\mathbf{\Psi e}+\boldsymbol{\mu})(\mathbf{\Lambda f}+\mathbf{\Psi e}+\boldsymbol{\mu})'\right\}.$$

*Proof.*

The then condition of the Theorem can be transformed into

$$\mathbf{\Sigma}=E\left\{\mathbf{\Lambda}[\mathbf{f}_{1i},\mathbf{f}_{2i}]\begin{bmatrix}\mathbf{f}_{1i}'\\ \mathbf{f}_{2i}'\end{bmatrix}\mathbf{\Lambda}'+\mathbf{\Psi}[\mathbf{e}_{1i},\mathbf{e}_{2i}]\begin{bmatrix}\mathbf{e}_{1i}'\\ \mathbf{e}_{2i}'\end{bmatrix}\mathbf{\Psi}+[\boldsymbol{\mu}_{1i},\boldsymbol{\mu}_{2i}]\begin{bmatrix}\boldsymbol{\mu}_{1i}'\\ \boldsymbol{\mu}_{2i}'\end{bmatrix}\right\}=E\left\{\mathbf{\Lambda ff}'\mathbf{\Lambda}'+\mathbf{\Psi ee}'\mathbf{\Psi}+\boldsymbol{\mu\mu}'\right\} \tag{4}$$

and

$$\mathbf{\Sigma}=\mathbf{\Lambda\Lambda}'+\mathbf{\Psi}^2+\begin{bmatrix}\mathbf{\Omega}_1 & \mathbf{0}\\ \mathbf{0} & \mathbf{\Omega}_2\end{bmatrix}=\mathbf{\Lambda\Lambda}'+\mathbf{\Psi}^2+\mathbf{\Omega}. \tag{5}$$

This completes the proof. □

For the model of essentially parallel measurements all non-diagonal elements of $\mathbf{\Sigma}$ are equal, so that a single non-diagonal element $\sigma_{\mathbf{X}_{t1},\mathbf{X}_{t2}}$ is considered in the following. Instead of two uncorrelated common factors, one may consider *q* uncorrelated common factors resulting in *q* item populations (each one based on one and the same common factor) and *q* subpopulations of individuals (each one working on only one subset of items). It follows from Theorem 1 that

$$\sigma_{\mathbf{X}_{t1},\mathbf{X}_{t2}} = E\left(\frac{1}{q}\sum_{i=1}^{q}\lambda \mathbf{f}_i\mathbf{f}_i'\lambda\right) = \lambda^2, \tag{6}$$

which implies that $\sigma_{\mathbf{X}_{t1},\mathbf{X}_{t2}}$ is the same as if all items were from one and the same factor even when the *q* factors are uncorrelated. Equation 6 remains unchanged for $q \to \infty$, which implies that $\sigma_{\mathbf{X}_{t1},\mathbf{X}_{t2}}$ is the same even when each individual responds to two items that are based on the same factor but on another factor than the two items of each other individual. Although it is rather unlikely that each individual responds to two items based on the same population whereas the item population is a different one for each individual, Equation 6 illustrates the limits of inferring factors from correlations. Even when each individual responds to items based from another item population based on factors that are uncorrelated with the factors that are the basis for the item populations of the other individuals, a one-factor model can be demonstrated, as long as all items the individual responds to are based on one and the same factor. A different interpretation of items by an individual does not affect the factor structure as long as it remains the same interpretation across all items for an individual.

Whereas it may be rather unlikely that the two items of each individual are exactly from the item population based on the same factor while there are *q* item populations based on *q* factors, the condition that each combination of two items from different item populations has an equal probability in the population of individuals is rather likely. Therefore, the latter condition is considered in the following.

## Items from different or identical populations across and within individuals

Starting from Equation 2 for two randomly selected items, it is assumed that 1/4 of the population of individuals work on each of the four possible item population combinations that are possible. The equal distribution of item population combinations on individuals results from an equal probability to select an item from the two item populations. For $\mathbf{\Lambda} = \mathbf{1}$, $\mathbf{\Psi} = diag(1 - \mathbf{\Lambda}\mathbf{\Lambda}')$, z-transformed measured variables $\mathbf{X}_{t1}$ and $\mathbf{X}_{t2}$, and $\mathbf{\Omega}_1 = \mathbf{\Omega}_2 = \mathbf{0}$, the correlation between the two items is

$$\rho_{\mathbf{X}_{t1},\mathbf{X}_{t2}} = E\left(\mathbf{X}_{t1}\mathbf{X}_{t2}'\right) = E\left(\frac{1}{4}\mathbf{f}_1\mathbf{f}_1' + \frac{1}{4}\mathbf{f}_1\mathbf{f}_2' + \frac{1}{4}\mathbf{f}_2\mathbf{f}_1' + \frac{1}{4}\mathbf{f}_2\mathbf{f}_2'\right) = \frac{1}{2}, \tag{7}$$

because $E(\mathbf{f}_1\mathbf{f}_2') = 0$. Thus, a substantial nonzero correlation between two items occurs in the total population comprising two subpopulations of individuals that worked on two items that were randomly selected from the same item population and two subpopulations that worked on two items selected from different item populations based on completely uncorrelated factors. When $\mathbf{\Lambda}$ and $\mathbf{\Psi}$ cannot be eliminated, and $\mathbf{\Omega}_1 = \mathbf{\Omega}_2 = \mathbf{0}$, the correlation between the two items in the total population of individuals is

$$\rho_{\mathbf{X}_{t1},\mathbf{X}_{t2}} = E\left(\frac{1}{4}\lambda\mathbf{f}_1\mathbf{f}_1'\lambda + \frac{1}{4}\lambda\mathbf{f}_2\mathbf{f}_2'\lambda\right) = \frac{1}{2}\lambda^2, \tag{8}$$

because of $E(\mathbf{f}_1\mathbf{f}_2') = 0$ and $E(\mathbf{e}_1\mathbf{e}_2') = 0$. For two items selected from three item populations based on uncorrelated factors similarly to Equation 1, nine combinations of item samples are possible ($\mathbf{X}_{s1}$, $\mathbf{X}_{s1}$; $\mathbf{X}_{s2}$, $\mathbf{X}_{s2}$; $\mathbf{X}_{s3}$, $\mathbf{X}_{s3}$; $\mathbf{X}_{s1}$, $\mathbf{X}_{s2}$; $\mathbf{X}_{s2}$, $\mathbf{X}_{s1}$; $\mathbf{X}_{s1}$, $\mathbf{X}_{s3}$; $\mathbf{X}_{s3}$, $\mathbf{X}_{s1}$; $\mathbf{X}_{s2}$, $\mathbf{X}_{s3}$; $\mathbf{X}_{s3}$, $\mathbf{X}_{s2}$). Three out of nine item combinations will be based on items from the same item population, so that assuming that each combination of items from different item populations occurs equally often in the total population of individuals results in

$$\rho_{\mathbf{X}_{t1},\mathbf{X}_{t2}} = E\left(\frac{1}{9}\lambda\mathbf{f}_1\mathbf{f}_1'\lambda + \frac{1}{9}\lambda\mathbf{f}_2\mathbf{f}_2'\lambda + \frac{1}{9}\lambda\mathbf{f}_3\mathbf{f}_3'\lambda\right) = \frac{1}{3}\lambda^2. \tag{9}$$

For $q^2$ equally sized independent subpopulations of individuals based on $q$ item populations based on $q$ common factors uncorrelated with the common factors of the other subpopulations of individuals, Equation 9 can be written as

$$\rho_{\mathbf{X}_{t1},\mathbf{X}_{t2}} = E\left(\frac{1}{q^2}\sum_{i=1}^{q}\sum_{j=1}^{q}\lambda\mathbf{f}_i\mathbf{f}_j'\lambda\right) = E\left(\frac{1}{q^2}\sum_{i=1}^{q}\lambda\mathbf{f}_i\mathbf{f}_i'\lambda + 2\frac{1}{q^2}\sum_{i=1}^{q}\sum_{j=i+1}^{q}\lambda\mathbf{f}_i\mathbf{f}_j'\lambda\right) = \frac{1}{q}\lambda^2, \tag{10}$$

because $E(\mathbf{f}_i\mathbf{f}_j') = 0$ for $i \neq j$. It should be noted that the $q^2$ subpopulations of individuals that occur for the items $\mathbf{X}_{t1}$ and $\mathbf{X}_{t2}$, will not be the same for any other pair of items. If, for example, a third item $\mathbf{X}_{t3}$ is presented and if the selection of items from the $q$ item populations into $\mathbf{X}_{t3}$ is again random, the $q^2$ equally sized independent subpopulations of individuals for $\rho_{\mathbf{X}_{t1},\mathbf{X}_{t3}}$ will not be the same as the $q^2$ subpopulations for $\rho_{\mathbf{X}_{t1},\mathbf{X}_{t2}}$. Nevertheless, assuming a model of parallel measurements will result in equal correlations between all $m$ presented items $\mathbf{X}_t = [\mathbf{X}_{t1} \ldots \mathbf{X}_{tm}]$. Accordingly, factor analysis of the correlation matrix of the $m$ items in the total population of individuals based on $q$ item populations results in

$$\mathbf{X}_t = \sqrt{\frac{1}{q}}\mathbf{\Lambda}\mathbf{f} + \sqrt{1-\frac{1}{q}}\mathbf{\Psi}\mathbf{e}. \tag{11}$$

It follows from Equation 8 that one factor models with moderate loadings may result from items selected from different item populations based on uncorrelated common factors measured in independent subpopulations of individuals. As different loadings in $\mathbf{\Lambda}$ may occur, it is impossible to ascertain the number of underlying subpopulations of individuals working on items based on different common factors, when the number and relative size of the respective subpopulations of individuals is not *a priori* known.

As can be seen from Figure 1A, a loading of about .50 may occur in factor analysis of the total population of individuals because the items load .50 on one dimension in all subpopulations of individuals, or because they load about .70 on two uncorrelated factors in different subpopulations of individuals, or about .80 on three uncorrelated factors in different subpopulations of individuals. Even a perfect loading on four uncorrelated factors in the subpopulations of individuals cannot be completely excluded, although this case is extremely unlikely. Nevertheless, when the factor loadings of the total factor analysis are greater than .71 it can be excluded at the population level that this loading results from two or more uncorrelated factors in different subpopulations of individuals.

If item mean variations are greater zero, the terms for the item means should be added to Equation 11. This yields

$$\mathbf{X}_{\mathrm{t}} = \sqrt{\frac{1}{q}}\boldsymbol{\Lambda}\mathbf{f} + \sqrt{1-\frac{1}{q}}\boldsymbol{\Psi}\mathbf{e} + [\boldsymbol{\mu}_1, \ldots, \boldsymbol{\mu}_q], \tag{12}$$

and

$$\boldsymbol{\Sigma}_{\mathrm{t}} = \frac{1}{q}\boldsymbol{\Lambda}\boldsymbol{\Lambda}' + \left(1-\frac{1}{q}\right)\boldsymbol{\Psi}^2 + \boldsymbol{\Omega}_t, \tag{13}$$

with $\boldsymbol{\Omega}_t = \begin{bmatrix} \boldsymbol{\Omega}_1 & \cdots & \mathbf{0} \\ \vdots & \ddots & \vdots \\ \mathbf{0} & \cdots & \boldsymbol{\Omega}_q \end{bmatrix}$ and $\boldsymbol{\Psi} = diag(1-\boldsymbol{\Lambda}\boldsymbol{\Lambda}')$, so that the item inter-correlations are

$$\mathbf{R}_{\mathrm{t}} = diag(1+\boldsymbol{\Omega}_t)^{-1/2}\boldsymbol{\Sigma}_{\mathrm{t}}\, diag(1+\boldsymbol{\Omega}_t)^{-1/2}. \tag{14}$$

Accordingly, a single non-diagonal element of the correlation matrix is

and

$$\rho_{\mathbf{X}_{\mathrm{t1}},\mathbf{X}_{\mathrm{t2}}} = \frac{1}{q(1+\varpi_{\mathrm{t1,t2}})}\lambda^2. \tag{15}$$

The relationship between common factor loadings and $q$ for $\varpi_{\mathrm{t1,t2}} = 1$ in the total population of individuals is given in Figure 1B. Of course, the size of resulting factor loadings for a given value of $q$ is smaller for $\varpi_{\mathrm{t1,t2}} = 1$ than for $\varpi_{\mathrm{t1,t2}} = 0$. Accordingly, when the factor loadings of the total factor analysis are greater than .50 it can be excluded at the population level that this loading results from two or more uncorrelated factors in different subpopulations of individuals.

(A)

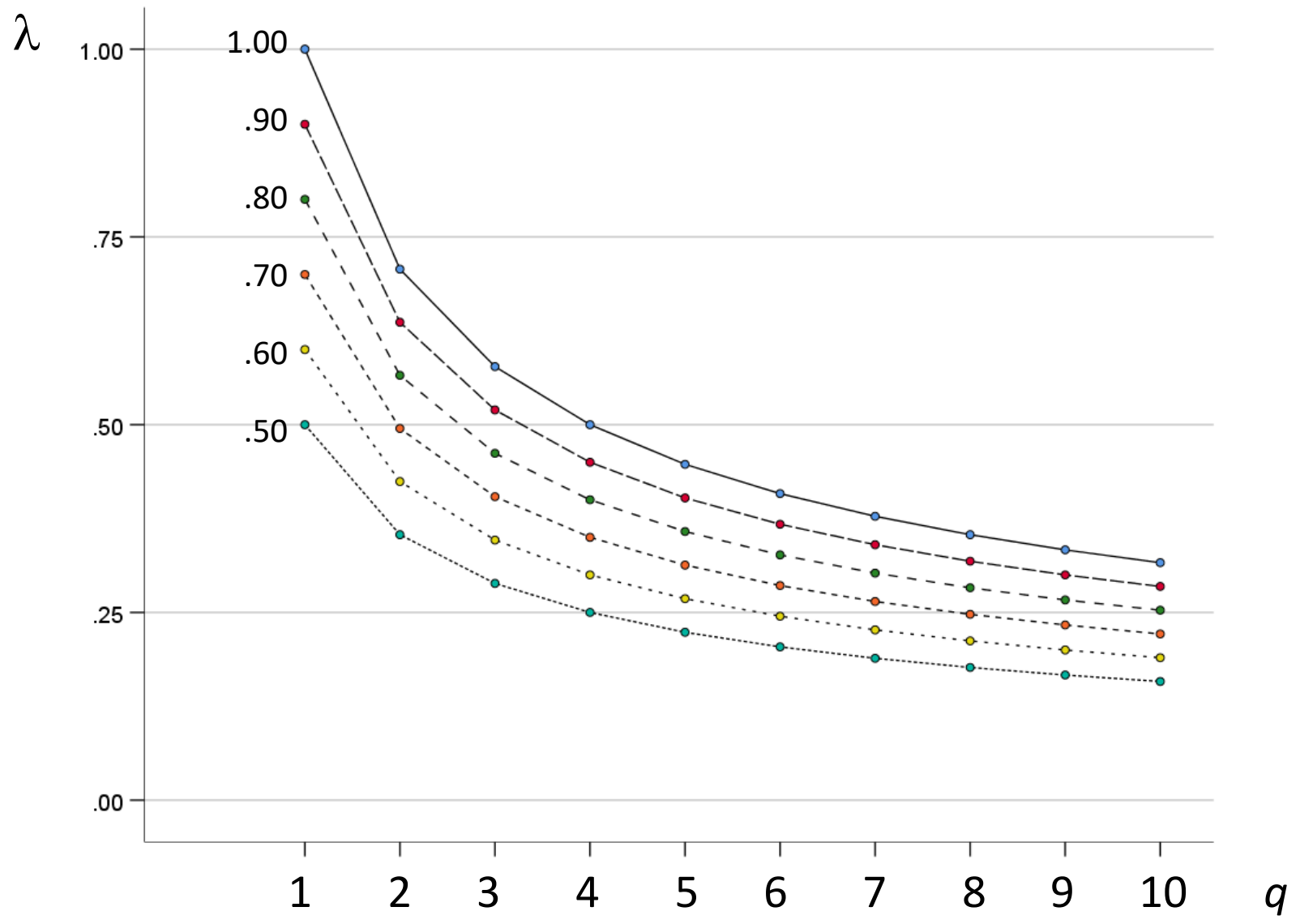

(B)

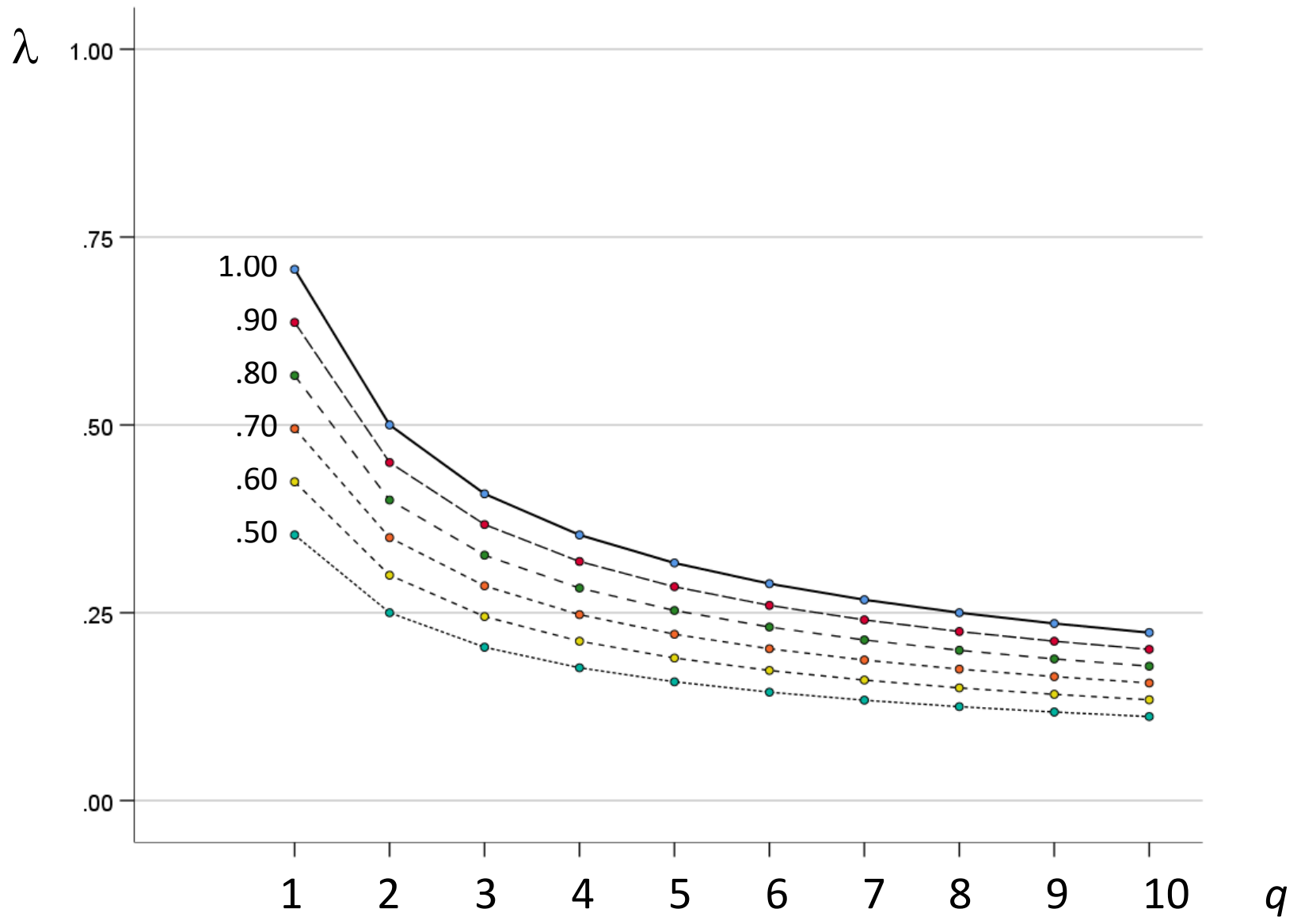

Figure 1. (A) Loadings of one-factor models in the total population of individuals based on $q^2$ subpopulations of individuals with $q$ uncorrelated factors for zero variance of item means; (B) loadings of one-factor models in the total population of individuals based on $q^2$ subpopulations of individuals with $q$ uncorrelated factors for unit-variance of item means.

The case of uncorrelated factors in the subpopulations is of special interest because it is the most critical possibility that might occur when researchers try to ascertain the dimensionality of a test, in which the participants worked on items that were randomly selected from different item populations that might be based on different factors. It is, however, rather likely that the common factors that determine the common variance of the items are not completely uncorrelated because the item populations that are used for a test might have a minimum of content similarity. When a non-zero correlation between the factors of different item populations is introduced, the respective equation for items selected from two item populations in two subpopulations of individuals is

$$\rho_{\mathbf{x}_{t1},\mathbf{x}_{t2}} = E\left(\frac{1}{4}\lambda\mathbf{f}_1\mathbf{f}_1'\lambda + \frac{1}{4}\lambda\mathbf{f}_2\mathbf{f}_2'\lambda + \frac{1}{4}\lambda\mathbf{f}_1\mathbf{f}_2'\lambda + \frac{1}{4}\lambda\mathbf{f}_2\mathbf{f}_1'\lambda\right) = \frac{1}{2}\lambda^2 + \frac{1}{2}\lambda^2\phi, \tag{16}$$

where $\phi$ is the non-zero correlation between $\mathbf{f}_1$ and $\mathbf{f}_2$. Since there is a non-zero correlation between the items in all subpopulations of individuals, there is a common factor across all subpopulations so that the uncorrelated factors that are specific for the subpopulations have a smaller effect on the correlation. In order to separate the part of the factor loadings that occur in all subpopulations of individuals Equation 16 can be written as

$$\rho_{\mathbf{x}_{t1},\mathbf{x}_{t2}} = \frac{1}{2}\lambda^2 - \frac{1}{2}\lambda^2\phi + \lambda^2\phi, \tag{17}$$

where $\lambda^2\phi$ represents the part of $\rho_{\mathbf{x}_{t1},\mathbf{x}_{t2}}$ that occurs in all subpopulations of individuals and $\frac{1}{2}\lambda^2 - \frac{1}{2}\lambda^2\phi$ is due to the remaining uncorrelated factors that are specific to the subpopulation of individuals and to the item population. For $q$ item populations the equation is

$$\rho_{\mathbf{X}_{t1},\mathbf{X}_{t2}} = \frac{1}{q}\lambda^2 - \frac{1}{q}\lambda^2\phi + \lambda^2\phi = \left(\frac{1}{q} - \frac{1}{q}\phi + \phi\right)\lambda^2. \tag{18}$$

The corresponding equation for factor analysis is

$$\mathbf{X}_t = \sqrt{\frac{1}{q} + \phi(1 - \frac{1}{q})}\mathbf{\Lambda f} + \sqrt{1 - \frac{1}{q} - \phi(1 - \frac{1}{q})}\mathbf{\Psi e}. \tag{19}$$

For $\phi \rightarrow 1$ Equation 19 approaches $\mathbf{X}_t = \mathbf{\Lambda f} + \mathbf{\Psi e}$, the conventional defining Equation of factor analysis (Harman, 1976), and for $q \rightarrow \infty$ Equation 19 approaches $\mathbf{X}_t = \sqrt{\phi}\mathbf{\Lambda f} + \sqrt{1-\phi}\mathbf{\Psi e}$. However, unless $q$ and $\phi$ are known, it will be impossible to disentangle the effect of subfactors that are specific to subpopulations of individuals and that represent different item populations from the effect of $\phi$, representing a common factor that occurs in all subpopulations. A factor analysis of the total population will result in a single loading for each column conflating all effects. The same holds for effects of a variation of item means in the item populations when $\mathbf{\Omega}_1 = \mathbf{\Omega}_2 > \mathbf{0}$, because Equation 19 is then

$$\mathbf{X}_t = \sqrt{\frac{1}{q} + \phi(1 - \frac{1}{q})}\mathbf{\Lambda f} + \sqrt{1 - \frac{1}{q} - \phi(1 - \frac{1}{q})}\mathbf{\Psi e} + \begin{bmatrix} \mathbf{\Omega}_1 & \mathbf{0} \\ \mathbf{0} & \mathbf{\Omega}_2 \end{bmatrix}. \tag{20}$$

The effect of item mean variability enhances the total item variance while the common variance of the items remains unchanged. The item means variation therefore leads to a reduction of the item inter-correlations which cannot disentangled from the effects of $\phi$ and $q$ on the item inter-correlations.

**On the detection of item populations representing uncorrelated factors**

The effect of blind and random item selection from item populations based on uncorrelated or correlated factors on the inter-item correlation has been shown. An important result from these demonstrations is that -when there is no variation in item means- it can be excluded that the items are based on $q > 1$ populations of uncorrelated factors when the item factor loadings are greater .71 (when there is substantial variation of item-means, even smaller factor loadings may indicate $q > 1$). However, the effect of sampling error has not been considered and it might be relevant to investigate whether selected items are from different item populations based on uncorrelated factors, even when the factor loadings are below .71. One way to get some indication whether between-subjects item heterogeneity occurred is to consider that the combination of items based on identical and different factors into the resulting item scores does not only affect the size of the correlation coefficient, but also the bivariate item distribution. In Figure 2A the scatterplot of a sample correlation of $r_{\mathbf{X}_{t1},\mathbf{X}_{t2}} = .451$ based on a very large sample of $n$ = 100,000 drawn from a population of individuals based on $q$ = 2 item populations with uncorrelated factors and without item-mean variation (with $\varpi_{t1,t2} = 0$ and $\rho_{\mathbf{X}_{t1},\mathbf{X}_{t2}} = \frac{1}{2}0.95^2 = .451$) can be compared to the scatterplot of $r_{\mathbf{X}_{t1},\mathbf{X}_{t2}} = .451$ based on items selected from a single item population representing one and the same factor (Figure 2B).

Normal distributed z-standardized random variables were generated by means of IBM SPSS Version 26. For the example correlation plotted in Figure 2A, the z-standardized random variables were entered into the two Equations 1 in order to generate the $q = 2$ item populations. Accordingly, the probability of an item for being randomly selected from population 1 or population 2 for each individual case was .50. For the example presented in Figure 2B all variables were selected from the same item population.

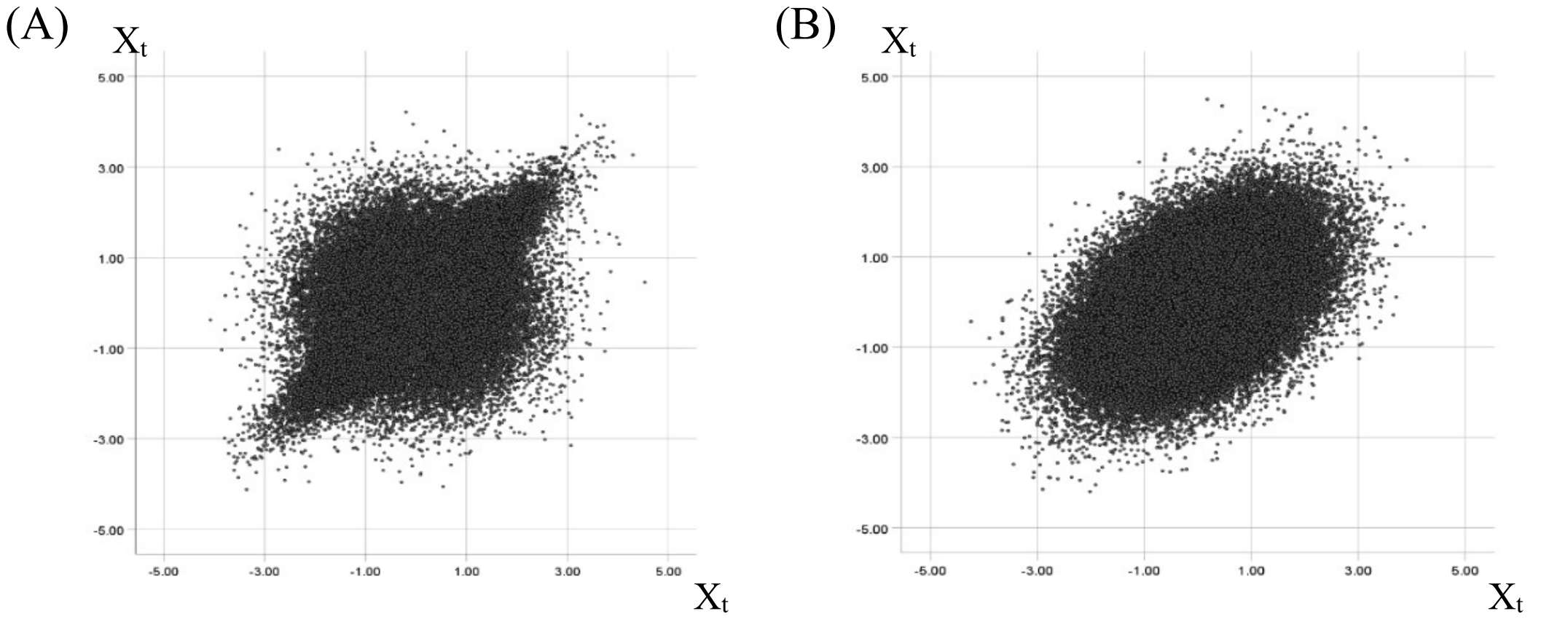

Figure 2. (A) Scatterplot of item scores from item populations based on two uncorrelated factors and $\varpi_{t1,t2} = 0$; (B) Scatterplot of item scores from items populations based on the same factor ($n$ = 100 000).

Obviously, one and the same correlation size is based on a rather different configuration of scores. Although the difference between the scatterplots will be less obvious for smaller correlations and smaller samples, the characteristic of the difference between the plots will probably remain the same.

An index of the degree of correlational heterogeneity can be based on the scores $\mathbf{c}_{1,2}$ with $E(\mathbf{c}_{1,2}\mathbf{c}_{1,2}') = \mathbf{I}$ of the two principal components that can be computed from

$$\begin{bmatrix} 1.00 & \rho_{\mathbf{X}_{t1},\mathbf{X}_{t2}} \\ \rho_{\mathbf{X}_{t1},\mathbf{X}_{t2}} & 1.00 \end{bmatrix} = \mathbf{\Sigma}_{t1,t2} = \mathbf{AA}'; \quad \begin{bmatrix} \mathbf{X}_{t1} \\ \mathbf{X}_{t2} \end{bmatrix} = \mathbf{X}_{t1,t2} = \mathbf{Ac}_{x1,x2}; \text{ and } (\mathbf{A}'\mathbf{A})^{-1}\mathbf{A}'\mathbf{X}_{t1,t2} = \mathbf{A}\begin{bmatrix} \mathbf{c}_{x1} \\ \mathbf{c}_{x2} \end{bmatrix}, \quad (21)$$

where $\mathbf{A}$ represents the 2 × 2 component loading matrix. The advantage of the component scores is that they are orthogonal and z-standardized so that the shapes of the scatterplots are more standardized. Figure 3 shows the scatterplots of the component scores for the sample used for the illustrative example in Figure 2. Obviously, there is a smaller number of points in the squares marked by the dotted lines in Figure 3A than in the squares marked by the dotted lines in Figure 3B.

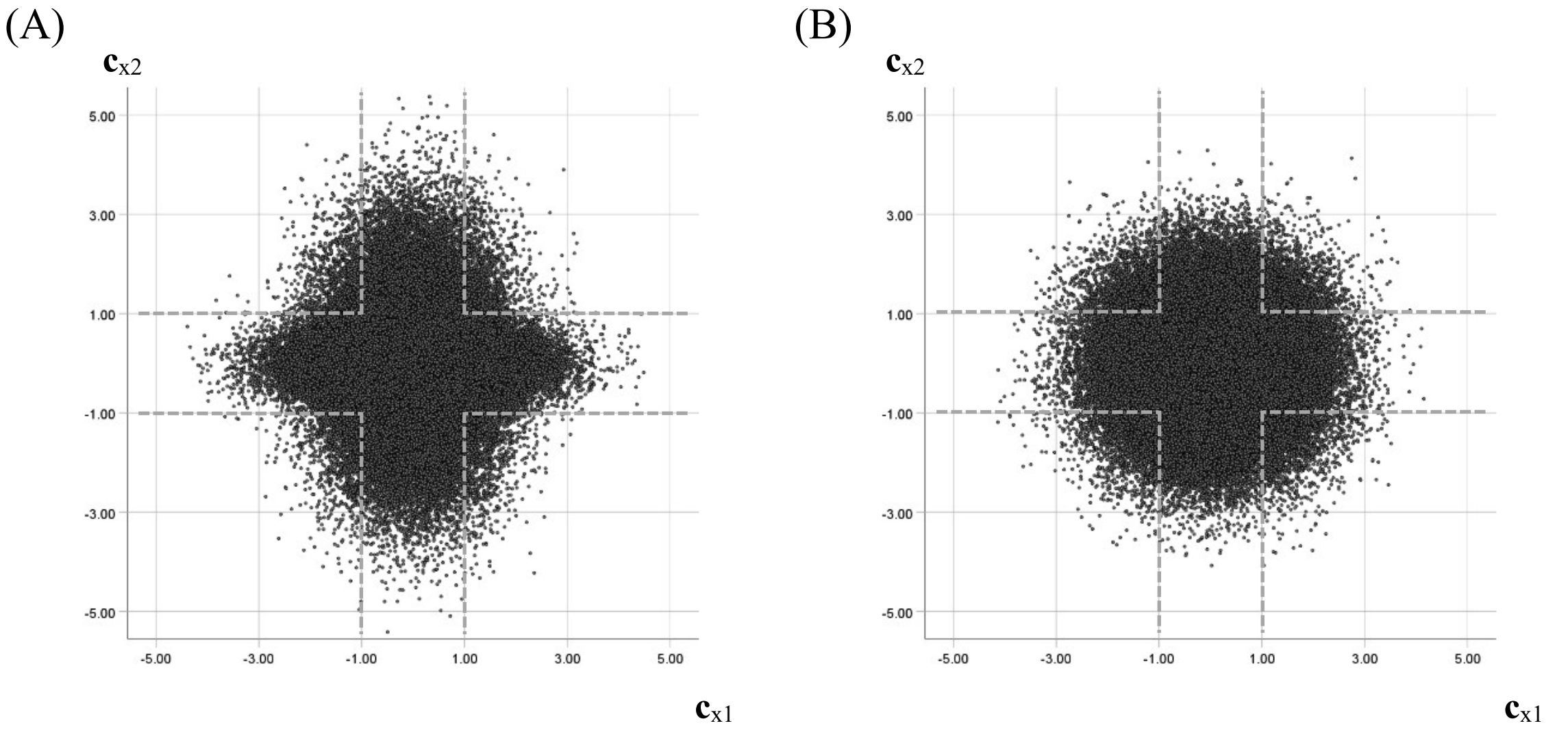

Figure 3. (A) Scatterplot of component scores from item populations based on two uncorrelated factors and $\varpi_{t1,t2} = 0$; (B) Scatterplot of component scores from items populations based on the same factor ($n$ = 100,000).

An index allowing for the detection of item populations from uncorrelated factors may be based on the different properties of the scatterplots of $\mathbf{c}_{x1}$ and $\mathbf{c}_{x2}$. Obviously, Figure 3B represents a bivariate normal distribution for $\rho = 0$ as the components are orthogonal, whereas the configuration of points of Figure 3A does not correspond to a bivariate normal distribution. As relevant differences between the distributions obviously occurred for larger scores, it was considered to use the larger scores for an index of deviations from the bivariate normal distribution. It is proposed here to compute the relative number of scores that are one standard deviation above or below the mean, i.e., scores with absolute values > 1 in both $\mathbf{c}_{x1}$ and $\mathbf{c}_{x2}$, which can be written as

$$v_{\mathrm{j}} = \begin{cases} v_j = 1 \ \mathit{if} \ \left|c_{\mathrm{x1j}}\right| > 1 \wedge \left|c_{\mathrm{x2j}}\right| > 1 \\ v_j = 0 \ \mathit{if} \ \left|c_{\mathrm{x1j}}\right| \le 1 \vee \left|c_{\mathrm{x2j}}\right| \le 1 \end{cases}, \text{for } j = 1 \text{ to } n. \quad (22)$$

The item population heterogeneity index $\kappa_{\mathrm{x}}$ is computed as

$$\kappa_{\mathrm{X}} = \frac{1}{n}\sum_{j=1}^{n} v_j. \quad (23)$$

For comparison with component scores based on $q$ = 1, z-standardized, normal distributed scores $\mathbf{Y}_{t1}$ and $\mathbf{Y}_{t2}$ are generated from Equation 7 for $q$ = 1 with the same sample size as the empirical data and with $\rho_{\mathbf{Y}_{t1},\mathbf{Y}_{t2}} = \rho_{\mathbf{X}_{t1},\mathbf{X}_{t2}}$. The resulting scores are submitted to principal component analysis and the number of component scores $\mathbf{c}_{y1}$ and $\mathbf{c}_{y2}$, with absolute values > 1 in both components is

$$w_j = \begin{cases} w_j = 1 \; if \left|c_{y1j}\right| > 1 \wedge \left|c_{y2j}\right| > 1 \\ w_j = 0 \; if \left|c_{y1j}\right| \le 1 \vee \left|c_{y2j}\right| \le 1 \end{cases}, \text{for } j = 1 \text{ to } n. \tag{24}$$

And the item population heterogeneity index for $q = 1$ is

$$\kappa_Y = \frac{1}{n}\sum_{j=1}^{n} w_j. \tag{25}$$

In the population, the relative number of points computed in Equation 18 corresponds to the probability of z-standardized scores of the standardized bivariate normal distribution (with $\rho$ = 0) for the condition $\left|c_{x1}\right| > 1 \wedge \left|c_{x2}\right| > 1$. The probability of a bivariate score smaller than one is $P(-\infty < c_{x1} < -1 \cap -\infty < c_{y1} < -1) = .0253$ (Statistics Online Computational Resource, SOCR), which represents the probability of scores in the left lower square segment in Figure 3 B. As the bivariate normal distribution for orthogonal scores is symmetric, all square segments have the same probability density so that the total population probability for the four squares is P = 0.1012 so that $\kappa_Y$ converges to this probability for $n \to \infty$.

In order to compare $\kappa_X$ and $\kappa_Y$ the latter should be based on a very large number of runs (at least 100) with the respective sample size and correlation. If $\kappa_X$ is smaller than the 5th percentile of the resulting $\kappa_Y$ distribution based on $q = 1$, one can assume that $\mathbf{X}_{t1}$ and $\mathbf{X}_{t2}$ are based on heterogeneous item populations ($q > 1$). For the example of scores presented in Figure 3 the mean of $\kappa_Y$ (for 100 runs) is .108, which is close to the population probability expected from the corresponding bivariate normal distribution. Moreover, $\kappa_X$ = .056 and the 5th percentile of $\kappa_Y$ is .099, indicating that the items $\mathbf{X}_{t1}$ and $\mathbf{X}_{t2}$ are based on heterogeneous item populations, which corresponds to the data generation based on $q = 2$ uncorrelated factors. However, for illustrative purposes, the sample size was extremely large so that effects of sampling error were probably negligible. In the following, a simulation study was performed in order to investigate the effects of $q$, $\rho_{\mathbf{X}_{t1},\mathbf{X}_{t2}}$, and sampling error on $\kappa_X$ and $\kappa_Y$.

**Simulation study on the detection of item populations representing uncorrelated factors**

A simulation study was performed in order to investigate the effect of sampling error on the detection rate of bivariate Pearson correlations based on $q > 1$ for $\varpi_{t1,t2} = 0$ by means of $\kappa_X$ and $\kappa_Y$. Data generation and statistical analysis was performed with IBM SPSS Version 26. Normal distributed, z-standardized random variables were computed by the method of Box and Muller (1958) from uniformly distributed numbers, which have been generated by the Mersenne twister integrated in SPSS. The random variables were entered into Equations 1 in order to generate the scores of observed variables, which are then correlated. Sample sizes were $n$ = 250, 500, and 1,000 cases, factor loadings of the generating factors were $\lambda$ = .70, .75, .80, .85, and .90, and $q$ = 1, 2, and 3. The $q$ = 1 condition, where all items are based on the same common factor, was included for comparison purposes. This results in 45 conditions for uncorrelated factors (3 sample sizes × 5 factor loadings × 3 different factors). For $q$ = 2 and 3 additional simulations were performed for the same parameters but with a factor inter-correlation of $\phi$ = .40, resulting in 30 additional conditions (3 sample sizes × 5 factor loadings × 2 different factors). For each condition 1,000 samples were drawn from the population in order to compute $\kappa_X$. In each sample, $\kappa_X$ was compared with the 5$^{th}$ percentile of $\kappa_Y$, which was based on 500 runs. A script for the simulation of detection rates is given in the Appendix.

The results for the uncorrelated factors are given in Table 1. For uncorrelated factors, inter-correlations ≥ .41 and sample sizes ≥ 500 cases, detection rates for $q$ = 2 were .90 or greater. For $q$ = 3, inter-correlations ≥ .27, and sample sizes ≥ 1,000 cases were necessary in order to get a detection rate of .95. For correlated factors ($\phi$ = .40), the detection rates were below .80 even when the analysis is based on 1,000 cases.

Table 1. Mean $\kappa_X$, standard deviation of $\kappa_X$ (in brackets), 5th percentile of $\kappa_Y$, detection rate: percent of $\kappa_X$ below 5th percentile of $\kappa_Y$ for uncorrelated factors ($\phi$ = .00)

| $q$ | $1/q$ $*\lambda^2$ = $\rho_{X_{t1},X_{t2}}$ | $n$ | $\kappa_X$ | | $\kappa_Y$ | 5th percentile of $\kappa_Y$ | detection rate: percent of $\kappa_X$ < 5th percentile of $\kappa_Y$ |
|---|---|---|---|---|---|---|---|
| | | 250 | .100 | (.016) | .100 | .074 | 4 |
| | 1/1*.70² =.49 | 500 | .100 | (.012) | .100 | .082 | 4 |
| | | 1,000 | .101 | (.008) | .101 | .087 | 4 |
| | | 250 | .100 | (.016) | .100 | .074 | 5 |
| | 1/1*.75² =.56 | 500 | .100 | (.011) | .100 | .082 | 5 |
| | | 1,000 | .101 | (.008) | .101 | .087 | 4 |
| | | 250 | .101 | (.016) | .100 | .074 | 4 |
| 1 | 1/1*.80² =.64 | 500 | .101 | (.012) | .100 | .082 | 5 |
| | | 1,000 | .101 | (.008) | .101 | .087 | 5 |
| | | 250 | .101 | (.017) | .100 | .074 | 5 |
| | 1/1*.85² =.72 | 500 | .101 | (.011) | .100 | .082 | 3 |
| | | 1,000 | .101 | (.008) | .101 | .087 | 5 |
| | | 250 | .100 | (.017) | .100 | .074 | 4 |
| | 1/1*.90² =.81 | 500 | .100 | (.012) | .100 | .082 | 4 |
| | | 1,000 | .101 | (.008) | .101 | .087 | 4 |
| | | 250 | .091 | (.016) | .100 | .074 | 11 |
| | 1/2*.70² =.25 | 500 | .091 | (.011) | .100 | .082 | 19 |
| | | 1,000 | .091 | (.008) | .101 | .087 | 28 |
| | | 250 | .088 | (.016) | .100 | .074 | 16 |
| | 1/2*.75² =.28 | 500 | .088 | (.011) | .100 | .082 | 26 |
| | | 1,000 | .088 | (.008) | .101 | .087 | 45 |
| | | 250 | .082 | (.016) | .100 | .074 | 26 |
| 2 | 1/2*.80² =.32 | 500 | .083 | (.011) | .100 | .082 | 41 |
| | | 1,000 | .083 | (.008) | .101 | .087 | 69 |
| | | 250 | .076 | (.016) | .100 | .074 | 44 |
| | 1/2*.85² =.36 | 500 | .076 | (.011) | .100 | .082 | 67 |
| | | 1,000 | .076 | (.008) | .101 | .087 | 91 |
| | | 250 | .066 | (.015) | .100 | .074 | 66 |
| | 1/2*.90² =.41 | 500 | .067 | (.011) | .100 | .081 | 90 |
| | | 1,000 | .066 | (.008) | .101 | .087 | 99 |
| | | 250 | .091 | (.016) | .100 | .074 | 10 |
| | 1/3*.70² =.16 | 500 | .092 | (.012) | .100 | .082 | 18 |
| | | 1,000 | .093 | (.008) | .101 | .087 | 21 |
| | | 250 | .089 | (.016) | .100 | .074 | 13 |
| | 1/3*.75² =.19 | 500 | .090 | (.011) | .100 | .081 | 22 |
| | | 1,000 | .090 | (.008) | .101 | .087 | 34 |
| | | 250 | .085 | (.016) | .100 | .074 | 19 |
| 3 | 1/3*.80² =.21 | 500 | .086 | (.011) | .100 | .082 | 33 |
| | | 1,000 | .086 | (.008) | .101 | .087 | 55 |
| | | 250 | .080 | (.016) | .100 | .074 | 30 |
| | 1/3*.85² =.24 | 500 | .081 | (.011) | .100 | .082 | 50 |
| | | 1,000 | .081 | (.008) | .101 | .087 | 79 |
| | | 250 | .075 | (.015) | .100 | .074 | 44 |
| | 1/3*.90² =.27 | 500 | .075 | (.011) | .100 | .082 | 70 |
| | | 1,000 | .075 | (.007) | .101 | .087 | 95 |

*Note.* The mean and standard deviation of $\kappa_X$ is based on 1,000 samples, the mean of $\kappa_Y$ is based on 500 runs in each of 1,000 samples.

Table 2. Mean and standard deviation (in brackets), 5th percentile of $\kappa_Y$ , detection rate: percent of $\kappa_X$ below 5th percentile of $\kappa_Y$ for correlated factors ( $\phi = .40$)

| $q$ | $\left(\frac{1-\phi+q\phi}{q}\right)\lambda^2 = \rho_{X_{t1},X_{t2}}$ | $n$ | $\kappa_X$ | | $\kappa_Y$ | 5th percentile of $\kappa_Y$ | detection rate: percent of $\kappa_X$ < 5th percentile of $\kappa_Y$ |
|---|---|---|---|---|---|---|---|
| | | 250 | .097 | (.016) | .100 | .074 | 6 |
| | (1-.40+2*.40)/2*.70² | 500 | .097 | (.011) | .100 | .082 | 8 |
| | =.34 | 1,000 | .096 | (.008) | .101 | .087 | 12 |
| | | 250 | .095 | (.016) | .100 | .074 | 9 |
| | (1-.40+2*.40)/2*.75² | 500 | .095 | (.012) | .100 | .082 | 11 |
| | =.39 | 1,000 | .095 | (.008) | .101 | .087 | 15 |
| | | 250 | .092 | (.016) | .100 | .074 | 10 |
| 2 | (1-.40+2*.40)/2*.80² | 500 | .092 | (.011) | .100 | .082 | 14 |
| | =.45 | 1,000 | .093 | (.008) | .101 | .087 | 22 |
| | | 250 | .089 | (.016) | .100 | .074 | 15 |
| | (1-.40+2*.40)/2*.85² | 500 | .088 | (.011) | .100 | .082 | 26 |
| | =.51 | 1,000 | .088 | (.008) | .101 | .087 | 41 |
| | | 250 | .082 | (.016) | .100 | .074 | 25 |
| | (1-.40+2*.40)/2*.90² | 500 | .082 | (.011) | .100 | .082 | 48 |
| | =.57 | 1,000 | .082 | (.008) | .101 | .087 | 76 |
| | | 250 | .097 | (.017) | .100 | .074 | 7 |
| | (1-.40+3*.40)/3*.70² | 500 | .097 | (.011) | .100 | .082 | 7 |
| | =.29 | 1,000 | .098 | (.008) | .101 | .087 | 9 |
| | | 250 | .096 | (.016) | .100 | .074 | 7 |
| | (1-.40+3*.40)/3*.75² | 500 | .096 | (.012) | .100 | .082 | 8 |
| | =.34 | 1,000 | .096 | (.008) | .101 | .087 | 13 |
| | | 250 | .092 | (.016) | .100 | .074 | 10 |
| 3 | (1-.40+3*.40)/3*.80² | 500 | .094 | (.011) | .100 | .082 | 12 |
| | =.38 | 1,000 | .094 | (.008) | .101 | .087 | 19 |
| | | 250 | .090 | (.016) | .100 | .074 | 14 |
| | (1-.40+3*.40)/3*.85² | 500 | .091 | (.011) | .100 | .082 | 19 |
| | =.43 | 1,000 | .091 | (.008) | .101 | .087 | 31 |
| | | 250 | .086 | (.016) | .100 | .074 | 18 |
| | (1-.40+3*.40)/3*.90² | 500 | .086 | (.011) | .100 | .082 | 30 |
| | =.49 | 1,000 | .086 | (.008) | .101 | .087 | 52 |

*Note.* The mean and standard deviation of $\kappa_X$ is based on 1,000 samples, the mean of $\kappa_Y$ is based on 500 runs in each of 1,000 samples.

## Discussion

Combinations of subpopulations of individuals responding to items from different item populations may occur when random item selection is performed. We investigated the effect of different subpopulations of individuals responding to items from different populations based on completely uncorrelated factors on the item inter-correlations and factor loadings in the total population of individuals. The analyses were based on the assumption that the model of essentially parallel measurements holds in each item population.

In a first step, the case that different subpopulations of individuals respond to items from different populations while each individual responds to items from the same population, was investigated. In this case, the number of subpopulations working on items from different populations does not affect the size of item inter-correlations. This implies that a one-factor model based on the item inter-correlations in the total population of individuals does not demonstrate that the randomly selected items are based on one and the same item population. A substantial bivariate correlation does not necessarily imply that there is a single factor explaining the common variance of two items because the responses of each individual to the two items can be determined by another factor for each individual. The only condition that must necessarily hold, is that both responses of a single individual are determined by the same factor. This corroborates the finding of Maraun and Heene (2015), who demonstrated that measurement invariance across populations does not necessarily imply equivalence of factors. Although the focus was on random item selection, this reasoning may be applied to any Pearson correlation, so that even the idea that a single common factor is the cause of the variation might be challenged. This result is a challenge for the validation of tests based on randomly selected items by means of factor analysis. It also implies that different individual interpretations of item content (i.e., *beta press*, Murray, 1938) do not affect a one-factor structure, when for each single individual the interpretation across all items remains the same. Therefore, person-item interactions, which are often investigated in the context of generalizability theory (Cronbach, Rajaratnam & Gleser, 1963; Medvedev, Krägeloh, Narayanan & Siegert, 2017), need not to affect a one-factor structure, when they are consistent for each individual across all items.

In a second step, the combination of items from different as well as identical item populations across and within individuals was investigated. In this case, some subpopulations of individuals respond to items from different populations based on uncorrelated factors and some subpopulations of individuals respond to items from one and the same population. It was shown that - under the assumption of the model of parallel measurements- a common factor loading larger than .71 in the total population of individuals indicates that the items are drawn from a single population. This implies that –when each individual responds to possibly different, randomly selected items from possibly different item populations– factor loadings greater .71 are necessary in order to conclude that the items have convergent validity across the total population of individuals.

Moreover, under the condition of subpopulations of individuals responding to items from the same as well as to items from different populations based on uncorrelated factors, a systematic deviation from the bivariate normal distribution of the items occurs in the total population of individuals. This deviation from the bivariate normal distribution was used as a basis for an indicator of the number of item subpopulations based on uncorrelated factors. This indicator was investigated in a simulation study. A detection rate of at least 90% was found for two item populations based on uncorrelated factors when the sample size was at least 500 and when the inter-item correlation was greater than .40. With a larger number of item populations and item populations based on correlated factors the detection rate was rather low. Nevertheless, the detection rate may be regarded as an indicator for the homogeneity of the convergent validity of items across subpopulations of individuals in the case of random item selection.

Although the focus was on items, the present results may also be of interest for the inter-correlation of scales. When data from different scales for the measurement of the same construct are pooled, the indicator proposed for the identification of the number of subpopulations of individuals working on variables from different populations may be used in order to investigate the homogeneity of the scales in the pooled sample. A general recommendation following from the present study is that, when bivariate correlations are interpreted, the effect of subpopulations of individuals working on different populations of variables should be considered.

## Appendix

### SPSS-Syntax for detection of the number of item populations:

```
* Encoding: windows-1252.

SET MXLOOPS=1000000 RNG=MT MDISPLAY=TABLES.
/* SET MTINDEX 20210419.

MATRIX.
/* - - - - - - - - - - - - - - .
/* Number of factors .
+  compute q=2.
/* Inter-factor correlation .
+  compute phi=.0.
/* Loading .
+  compute L=.9.
/* Variance of means .
+  compute omega=0.
/* Sample size .
+  compute n=250.
/* Number of Monte-Carlo-Samples .
+  compute nsamples=1000.
/* Number of runs for kappa(y) .
+  compute nruns=500.
/* Bootstrap (=1) or normal distribution (=0) for kappa(y) .
+  compute boot=0.
/* - - - - - - - - - - - - - - .
+  do if phi=0 and q>1.
+     compute phiq=ident(q).
+     compute psi=mdiag(sqrt(1-diag(L*phiq*t(L)))).
+     else if phi ne 0 and q > 1.
+     compute phiq=mdiag(make(q,1,(1-phi)))+phi.
+     compute psi=mdiag(sqrt(1-diag(L*phiq*t(L)))).
+     else if q=1.
+     compute phiq=1.
+     compute psi=(1-L**2)**.5.
+  end if.
+  compute result=make(1,4,0).
+  loop ii=1 to nsamples.
+     compute fs=((-2&*ln(uniform(n,q)))&**0.5)&*(cos(4*arsin(1)&*uniform(n,q)))*chol(phiq).
```

```
+     compute sample=make(n,2,0).
+     loop jj=1 to 2.
+        compute es=((-2&*ln(uniform(n,q)))&**0.5)&*(cos(4*arsin(1)&*uniform(n,q))).
+        do if omega>0.
+           compute mus=((-2&*ln(uniform(n,q)))&**0.5)&*(cos(4*arsin(1)&*uniform(n,q)))*omega**0.5.
+           else.
+           compute mus=0.
+        end if.
+        compute data=fs*t(L)+es*psi+mus.
+        compute rndm=uniform(n,q).
+        compute rndm=rsum(data&*(rndm=(rmax(rndm)*make(1,q,1)))).
+        compute sample(:,jj)=rndm.
+     end loop.
+     compute xc=(make(1,n,1)*sample)/n.
+     compute s=(t(sample)*sample-n*(t(xc)*(xc)))/(n-1).
+     compute ivar=inv(mdiag(sqrt(diag(s)))).
+     compute R=ivar*S*ivar.
+     compute xv=sample-make(n,1,1)*xc.
+     compute zv=xv*ivar.
+     call eigen(R,V,ew).
+     compute scores=zv*V*mdiag(ew&**(-.5)).
+     compute kappax=csum(rmin(abs(scores)>1))/n.
+     compute zsample={sample(:,1);sample(:,2)}.
+     compute msample=csum(zsample)/nrow(zsample).
+     compute ssample=(cssq(zsample-msample)/(nrow(zsample)-1))**.5.
+     compute zsample=(zsample-msample)/ssample.
+     compute kappay=0.
+     loop kk=1 to nruns.
+        do if boot=0.
/* Sampling from normal distribution .
+           compute rsample=((-2&*ln(uniform(n,2)))&**0.5)&*(cos(4*arsin(1)&*uniform(n,2)))*chol(R).
+           else.
/* Sampling from observed data (Bootstrap) .
+           compute rsample=({zsample((trunc(2*n*uniform(n,1)+1)),1),zsample((trunc(2*n*uniform(n,1)+1)),1)})*chol(R).
+        end if.
+        compute xc=(make(1,n,1)*rsample)/n.
+        compute s=(t(rsample)*rsample-n*(t(xc)*(xc)))/(n-1).
+        compute ivar=inv(mdiag(sqrt(diag(s)))).
+        compute Rrs=ivar*S*ivar.
+        compute xv=rsample-make(n,1,1)*xc.
+        compute zv=xv*ivar.
+        call eigen(Rrs,V,ew).
+        compute rscores=zv*V*mdiag(ew&**(-.5)).
+        compute kappay={kappay;(csum(rmin(abs(rscores)>1))/n)}.
```

```
+     end loop.
+     compute kappay=kappay(2:nrow(kappay)).
+     compute p05ky=csum((trunc(nrow(kappay)*.05)=grade(kappay))&*kappay).
+     compute mky=csum(kappay)/nrow(kappay).
+     compute detect=kappax<p05ky.
+     compute result={result;({kappax,mky,p05ky,detect})}.
/*+     compute out={nsamples,nruns,boot,phi,q,L,omega,n,kappax,mky,p05ky,detect}.
/*+     save out /outfile=* /variables=nsamples,nruns,boot,phi,q,L,omega,n,kappax,mky,p05ky,detect /* kappax<p05ky */.
+  end loop.
+  compute result=result(2:nrow(result),:).
+  compute mr=csum(result)/nrow(result).
+  compute skx=(cssq(result(:,1)-mr(1,1))/(nrow(result)-1))**.5.
+  print {nsamples,nruns,boot,phi,q,L,omega,n,mr(1,1),skx,mr(1,2:3),(100*mr(1,4))}
      /clabels=nsamples,nruns,boot,phi,q,L,omega,n,kappax,skx,kappay,p05ky,"proportion of kappax < p05ky"
      /format=F7.3.
END MATRIX.
```